\input mn

\pageoffset{-2.5pc}{0pc}

\pagerange{1}
\pubyear{1997}
\volume{000}

\begintopmatter  
\title{Metallicity effects on the chromospheric activity--age relation 
    for late-type dwarfs}
\author{H. J. Rocha-Pinto and W. J. Maciel}
\affiliation{Instituto Astron\^omico e Geof\'{\i}sico, Av. Miguel 
    Stefano 4200, 04301-904 S\~ao Paulo, Brazil}
	       
\shortauthor{H. J. Rocha-Pinto and W. J. Maciel}
\shorttitle{The chromospheric activity--age relation}


\acceptedline{Accepted. Received.}

\abstract {We show that there is a relationship between the age excess,
    defined as the difference between the stellar isochrone and 
    chromospheric ages, and the metallicity as measured by the index 
    [Fe/H] for late-type dwarfs. The chromospheric age tends to be 
    lower than the isochrone age for metal-poor stars, and the 
    opposite occurs for metal-rich objects. We suggest that this 
    could be an effect of neglecting the metallicity dependence of 
    the calibrated chromospheric emission--age relation. We propose 
    a correction to account for this dependence. We also investigate 
    the metallicity distributions of these stars, and show that there 
    are distinct trends according to the chromospheric activity level. 
    Inactive stars have a metallicity distribution which resembles the 
    metallicity distribution of solar neighbourhood stars (Rocha-Pinto 
    \& Maciel 1996), while active stars appear to be concentrated in 
    an activity strip on the $\log R'_{\rm HK} \times{\rm [Fe/H]}$ 
    diagram. We provide some explanations for these trends, and show 
    that the chromospheric emission--age relation probably has 
    different slopes on the two sides of the Vaughan-Preston gap.} 
. 

\maketitle

\section{Introduction}

The usual method for estimating ages of field stars consists in 
comparing the star position, in a $M_V\times\log T_{\rm eff}$ diagram, 
with respect to a grid of theoretical isochrones. This procedure is 
affected by errors in both coordinates, although the uncertainties
in the determination of the bolometric corrections and effective 
temperatures are particularly important.

An alternate method, which seems to be very promising for estimating 
ages in late-type dwarfs, makes use of the chromospheric emission (CE)
in these stars. From the pioneering work by Wilson (1963), there are 
many evidences in the literature that the stellar chromospheric 
activity can be associated with the stellar age (e.g., Skumanich 
1972; Barry, Cromwell \& Hege 1987; Eggen 1990; Soderblom, Duncan \& 
Johnson 1991). According to these investigations, young stars show CE 
levels systematically higher than the older stars. 

Recently, Henry et al. (1996, hereafter HSDB) published the results of 
an extensive survey of CE in southern hemisphere G dwarfs. These data, 
together with the data previously published by Soderblom (1985), make 
possible the study of the CE in a large number of stars with varying 
ages and chemical compositions. This work attempts to discuss the 
estimate of stellar ages using chromospheric indices in stars with 
different chemical compositions.  In section 2, we make a comparison 
between isochrone and chromospheric ages, and show that the 
chromospheric ages present systematic deviations related to the 
isochrone ages, as a function of the metallicity [Fe/H]. In section~3, 
we study the differences in the metallicity distributions of the active 
and inactive stars, and show that the active stars seem to be located 
in a strip in the CE--metallicity diagram. In section~4 we give some 
explanations for this strip and for other trends in the diagram. 
Finally, the CE--age relation on both sides of the Vaughan-Preston 
gap is treated in section 5.

\section{A comparison of isochrone and chromospheric ages}

Soderblom et al. (1991) demonstrated that the CE level, as measured by 
the index $\log R'_{\rm HK}$, is directly related to the stellar age. 
This relation is not a statistical one, but a deterministic relation, 
the form of which has not been well determined. Due to the scatter 
in the data, Soderblom et al. decided not to choose between a simple 
power-law calibration, which would produce many very young stars, 
and a more complicated calibration which would preserve the constancy
of the star formation rate.

Also, they argued that the relation between CE and stellar age seems 
to be independent of [Fe/H], as the Hyades and Coma clusters show 
the same CE levels and have the same age, although with different 
chemical compositions. In spite of that, the authors acknowledge
that low metallicity stars should have a different CE--age 
relation, since  in these stars the Ca {\sevenrm II} H and K lines 
are intrinsically shallower than in solar metallicity stars. In this 
sense, old metal-poor stars would resemble very young stars due to a 
higher $\log R'_{\rm HK}$.

To investigate the metallicity dependence of the chromospheric
activity-age relation, we have used the recent database by 
Edvardsson et al. (1993; Edv93), which comprises mainly late F and 
early G dwarfs. Since the surveys of HSDB and Soderblom (1985) include 
mainly G dwarfs, only a fraction of the stars in the Edv93 sample 
have measured chromospheric indices. There are 44 stars in common 
between these works, which we will call Sample~A. For these stars,
we have taken isochrone ages from Edv93 and estimated chromospheric
ages using Eq.~3 of Soderblom et al. (1991), which can be written as
$$\log t_{ce} = -1.50 \log R'_{\rm HK} + 2.25 \eqno\stepeq$$
where $t_{ce}$ is in yr. This is equivalent to $R'_{\rm HK} \propto 
t^{-2/3}$, and corresponds to the best power-law fit to their data. 
It should be noted that the use of any of the relations suggested by
Soderblom et al. (1991) would produce essentially the same results.
In Fig. 1 we show the difference between the isochrone ($t_{is}$) and 
chromospheric ($t_{ce}$) ages, or {\it age excess}, as a function  
of the metallicity [Fe/H],
$$\Delta(\log t) = \log t_{is} - \log t_{ce} .\eqno\stepeq$$
\beginfigure*{1}
  \vskip 12 cm
  \caption{{\bf Figure 1.} Chromospheric age excesses as a function of  
  [Fe/H]. a) Sample A (44 stars), using Edv93 ages. b) Sample A using 
  Ng \&  Bertelli (1998) ages. c) Sample B (471 stars). Symbols are
  as follows: very active stars (plus signs), active stars (open 
  circles), inactive stars (filled circles), and very inactive stars
  (crosses).}
\endfigure

We have used open circles for active stars ($\log R'_{\rm HK}\ge 
-4.75$), and filled circles for inactive stars ($\log R'_{\rm HK}< 
-4.75$, cf. Vaughan \& Preston 1980). Note that, 
except for a few isolated points, there is a clearcut relation 
between the age excesses and [Fe/H], especially for the
inactive stars, in the sense that the chromospheric ages are 
substantially lower than the isochrone ages for metal-poor objects, 
while the opposite occurs for metal-rich stars. The age excess is 
minimal for metallicities around solar. 

Recently, Ng \& Bertelli (1998) revised the Edv93 ages, incorporating 
new updated isochrones and absolute magnitudes from Hipparcos 
parallaxes. Figure~1b takes into account these revised ages for the 
stars in Sample~A. Note that the scatter is somewhat lower, and 
the general trend of Figure~1a is confirmed. The curve in Figs.~1a and 
1b corresponds to a third-order polynomial fitted to the inactive
stars with ages by Ng \& Bertelli (1998).

Besides Sample~A, we have built an independent sample with 730 stars
(hereafter Sample~B) by the intersection of the surveys from HSDB and 
Soderblom (1985) with the {\it uvby} catalogues of Olsen (1983, 1993, 
1994). The metallicity of each star in Sample B was estimated from 
the calibrations of Schuster \& Nissen (1989). The metallicities 
for 3 stars redder than $(b-y)=0.599$ were given by the calibration 
of Olsen (1984) for K2-M2 stars. In order to obtain the colours
$\delta m_1$ and $\delta c_1$, we have adopted the standard curves
$(b-y)\times m_1$ and $(b-y)\times c_1$ given by Crawford (1975) for 
late F and G0 dwarfs, and by Olsen (1984) for mid and late G dwarfs. 
The zero-age main sequence (ZAMS) curve $(b-y)\times c_1$ was 
corrected as in Edv93 to allow for a dependence on [Fe/H]. Finally, 
$\log T_{\rm eff}$ and $\Delta M_V$ (the absolute magnitude deviation
above the ZAMS) were calculated by the calibrations presented by 
Olsen (1984). Since all stars in Sample B are expected to be 
located within 50 pc around the Sun, no correction for reddening was 
applied to the photometric indices.

In order to obtain the ages, theoretical isochrones by VandenBerg 
(1985) were used. He presents grids of isochrones for metallicities 
[Fe/H] = 0.0, $-0.23$, $-0.46$, $-0.76$ and $-1.00$, and He
abundance $Y=0.25$. Following the general practice in isochrone 
age determinations (e.g., Twarog 1980; Soderblom et al. 1991; Edv93), 
we transformed his isochrones in the plane $M_V\times\log T_{\rm eff}$ 
to the plane $\Delta M_V\times\log T_{\rm eff}$, using tabulated ZAMS 
curves given by VandenBerg (1985). Also, we have shifted the resulting 
isochrones by $\delta\log T_{\rm eff}=-0.009$ as VandenBerg's models 
seem to be slightly too hot for disk stars, according to Edv93. We 
plotted each star from Sample B in these five sets of isochrone grids. 
An age for the star was estimated by graphic interpolation in each grid.
A first- or second-order polynomial was fitted to these ages for 
each star. The stellar age was found by applying the photometrically 
calculated [Fe/H] to the polynomial. We have estimated ages only for 
the stars which had ages in at least two set of grids, amounting
to 471 stars.

Table~1 shows the 730 stars of Sample B\note{$^\star$}{Table 1 is only 
available in electronic form.}. In the first column we give the 
HD number. The next columns list, respectively,  $\log R'_{\rm HK}$, 
[Fe/H], $\log T_{\rm eff}$, $\Delta M_V$, the isochrone logarithmic 
age, the chromospheric logarithmic age, and the age excess. The 
average errors in $\log T_{\rm eff}$ and $\Delta M_V$ can be estimated 
from the calibrations we have used as 
$\sigma\left(\log T_{\rm eff}\right)\sim\pm 0.011$ and 
$\sigma\left(\Delta M_V\right)\sim 0.29\ {\rm mag}$. These errors 
imply an age uncertainty of 1.5 to 3 Gyr for the hotter and cooler
stars, respectively. An error of about 0.16 dex in the metallicity 
calibration further increases the age uncertainty by about 1 Gyr. 

In Figure 1c, we show the age excesses for sample B. Besides the 
symbols used in the previous panels, we use plus signs for very 
active stars ($\log R'_{\rm HK}> -4.20$) and crosses for very 
inactive stars ($\log R'_{\rm HK}< -5.10$). The same general trend
arises in this figure: metal-poor stars show lower chromospheric 
ages [$\Delta(\log t) > 0$], while metal-rich stars show higher 
chromospheric ages [$\Delta(\log t) < 0$].  The curve fitted to 
sample~A is included for comparison. It can be seen that a number 
of stars deviates from the general trend, especially active and very 
active stars, which can also be observed  in Figure~1a. On the other 
hand, almost all the stars which follow the observed 
trend  between age excess and [Fe/H] have $\log R'_{\rm HK}< -4.75$, 
and are inactive or very inactive stars. In Figure 1b the same
trend can be observed, although the behaviour of the active
stars is not as clear. In the next section, we will try to understand 
why these two classes of stars, which were defined according to the 
chromospheric activity level, present different trends regarding 
their age excesses.

The data from sample A is more accurate than the data from sample B, 
since the photometric calibrations for F and early G dwarfs are 
better defined than for late G dwarfs. Moreover, we used a 
spectroscopic metallicity in Sample~A, and a photometric one in 
Sample~B. This is reflected in the higher scatter observed in 
Figure~1c as compared with Figures~1a,b. Note that the scatter
observed in Fig.~1 is probably real, as the chromospheric ages 
depend on the CE index $R'_{\rm HK}$ which is expected to vary
in small time scales, as a result of the stellar magnetic cycle. 
For example, for the sun a variation in the range 
$-5.10 \leq \log R'_{\rm HK} \leq -4.75$ is observed (cf. HSDB). 
Therefore, we decided to use sample~B only as an independent check 
of the results of Figures~1ab, and derived a correction to the 
chromospheric age on the basis of the results shown in figures~1a,b. 
In this case, the {\it corrected chromospheric age} can be written as
$$\log t_{ce}({\rm corrected}) = \log t_{ce} + \Delta(\log t) 
    \eqno\stepeq$$
where
$$\Delta(\log t) = -0.193 - 1.382 {\rm [Fe/H]} - 0.213 {\rm [Fe/H]}^2$$
$$\hskip 3.0 true cm   + 0.270 {\rm [Fe/H]}^3, \eqno\stepeq$$
which corresponds to the polynomial fit in Figure~1b. Note that this 
correction should be added to the logarithm of the chromospheric 
age, given in Gyr. This correction can be applied to chromospheric
ages derived from published CE--age calibrations (Soderblom et al. 
1991; Donahue 1993, as quoted by  HSDB) for $\log R'_{\rm HK}< -4.75$
in the range \hbox{$-1.2 < {\rm [Fe/H]} < +0.4$.}

We checked in the catalogue of Cayrel de Strobel et al. (1997) 
for the metallicities of the stars used by Soderblom et al. (1991) 
for the building of the CE--age relation. The mean metallicity of 
these stars is $-0.05$ dex, with a standard deviation of 0.16 dex. 
Referring to Figures 1a and 1b, we can see that for 
${\rm [Fe/H]} = -0.05$ we have $\Delta(\log t) \simeq 0$, which 
reinforces our conclusion that the published  CE--age calibrations 
are valid for solar metallicity stars only.
\beginfigure{2}
  \vskip 6.0 cm
  \caption{{\bf Figure 2.} CE index as a function of the Ca 
   abundances for the stars in Sample A. The values of [Ca/H] are 
   from  Edv93. Open circles: active stars; filled circles: 
   inactive stars.}
\endfigure
Another evidence favouring our conclusions is presented in Figure~2, 
where we plot the chromospheric emission index for the stars of 
sample~A as a function of the calcium abundance [Ca/H] taken
from Edv93. Again for the inactive stars, the figure shows that the 
calcium-rich stars, which are expected to be also metal-rich and young,
have generally lower CE indices. This can be explained by the fact
that these stars have stronger Ca features,  which leads to a lower
$R'_{\rm HK}$ and correspondingly higher chromospheric ages, as
observed in Figure~1a. Inversely, calcium-poor stars show higher 
chromospheric indices, which is mainly due to their weaker Ca 
spectral features.

\section{Metallicity distributions according to CE level}

Vaughan \& Preston (1980) noted that solar neighbourhood stars could 
be divided into two populations, namely active and inactive stars, 
herafter referred to as AS and IS, respectively. The separation 
between these populations is made by the so-called Vaughan-Preston gap, 
located around $\log R'_{\rm HK} = -4.75$, which is a region of 
intermediate activity containing very few stars. Recently, HSDB showed
that the Vaughan-Preston gap is a transition zone instead of a zone 
of avoidance of stars. Moreover, they showed that two additional 
populations seem to exist: the very active stars (VAS; 
$\log R'_{\rm HK} > -4.20$) and the very inactive stars (VIS; 
$\log R'_{\rm HK} < -5.10)$.

It is interesting to see whether any trends in the chemical
composition of these 4 groups exist. At a first glance, on the 
basis of the existence of a deterministic CE--age relation (Soderblom 
et al. 1991), and of the age--metallicity relation, we would 
expect a CE--metallicity relation to exist. The dependence on 
[Fe/H] of the CE--age relation, discussed in the previous section, 
could hinder or mask such a CE--metallicity relation, and our results 
could in principle show whether or not that dependence is real.

Figure 3 shows the ${\rm [Fe/H]} \times \log R'_{\rm HK} $ diagram 
for the 730 stars of Sample B. The vertical dashed lines 
separate the four populations according to their activity level. 
Contrary to the argument presented in 
the previous paragraph, there seems to be no significant  
CE--metallicity relation. Moreover, it is clear that the metallicity 
distribution of the four groups are very different from each other. 
This is shown more clearly in Figure~4, where we compare their 
metallicity distributions with the solar neighbourhood metallicity 
distribution (Rocha-Pinto \& Maciel 1996).
\beginfigure{3}
  \vskip 6.0 cm
  \caption{{\bf Figure 3.} ${\rm [Fe/H]}\times \log R'_{\rm HK}$ 
    diagram for the 730 stars of Sample B. The vertical dashed 
    lines separate the four populations. The solid lines mark an 
    apparent \lq activity strip\rq\  where the AS and VAS are mainly 
    located.}
\endfigure
\beginfigure{4}
  \vskip 6.0 cm
  \caption{{\bf Figure 4.} Metallicity distributions of the four 
     populations compared to the solar neighbourhood metallicity
     distribution from Rocha-Pinto \& Maciel (1996).}
\endfigure
It can be seen from Figure~4 that the metallicity distribution of the
IS is very similar to the solar neighbourhood standard distribution, 
suggesting that the IS sample represents very well the stars in our 
vicinity. This is not surprising, since about 70\% of the stars
in the solar neighbourhood are inactive (cf. HSDB). The other 
metallicity distributions show some deviations
from the standard distribution, as follows: (i) The paucity of 
metal-poor stars in the VIS group is peculiar, as these 
should be the oldest stars in the Galaxy, according to their low CE 
level, and the oldest stars are supposed to be metal-poor 
if the chemical evolution theory is correct. (ii) The lack of 
metal-poor stars is even more pronounced in the AS group, and the 
transition from the IS to the AS at the Vaughan-Preston gap suggests 
an abrupt change in the metallicity distribution of these stars. 
The AS have metallicities in a narrow range from about $-0.35$ dex 
to +0.15 dex, with an average around $-0.15$ dex. It is interesting 
to note that the majority of stars with ${\rm [Fe/H]} > +0.20$, 
which should be very young according to the chemical evolution, 
are not AS (young stars according to published CE--age 
relations), but appear amongst the VIS and the IS (old stars 
according to CE level). (iii) There are very few VAS, but their 
metallicities seem to be lower than the metallicities of the AS. 
From Fig.~3 it can be seen that the VAS, together with 
the AS, appear to be located mainly in a well-defined \lq activity 
strip\rq\ on the ${\rm [Fe/H]} \times \log R'_{\rm HK} $ diagram, 
which we have marked by solid lines in this figure. 

The nature of the VIS and the VAS was already treated by HSDB. The VIS 
are probably stars experiencing a phase similar to the Maunder-Minimum 
observed in the sun. The paucity of metal-poor stars in the VIS group 
reinforces this conclusion, by ruling out the hypothesis of old ages for 
these stars. 

The VAS seem to be formed mainly by close binaries. HSDB have confirmed 
that many of the VAS in their sample are known to be RS CVn or W UMa 
binaries, and have been detected by ultraviolet and x-ray satellites. 
In fact, a CE--age relation is not supposed to be valid for binaries 
(at least close binaries), since these stars can keep high CE levels, 
even at advanced ages, by synchronizing their rotation 
with the orbital motion (Barrado et al. 1994, Montes et al. 1996). 
It is possible that not all VAS are close binaries, but some of them 
can be instead very young stars. It is significant that the BY Draconis 
systems are composed by either close binaries and very young field 
stars (Fekel, Moffett \& Henry 1986), and that also the only 
difference between BY Dra and RS CVn stars is that BY Dra have later
spectral types (Eker 1992). However, taking into account their 
metallicity distribution, which shows no stars with 
${\rm [Fe/H]}>-0.10$ dex, it would appear that the close binaries 
hypothesis is more appealing, as very young stars should be metal-rich. 

Nevertheless, the situation is not so clear. The low [Fe/H] content 
of the VAS could be partially responsible for the high 
$\log R'_{\rm HK}$ indices in these stars, by the reasons explained 
in Section 2. On the other hand, it is also known that a high 
chromospheric activity can decrease the $m_1$ index by filling up 
the cores of the metallic lines, simulating a lower [Fe/H] 
content (Giampapa, Worden \& Gilliam 1979; Basri, Wilcots \& Stout 
1989). Gim\'enez et al. (1991) have shown that the $m_1$ deficiency 
in active systems like RS CVn stars is severe and can substantially 
affect the values of the photometrically derived stellar parameters. 
It should be recalled that even the Sun, which is an inactive star, 
would look 35\% more metal-poor if observed in a region of activity 
(Giampapa et al. 1979). Thus, it is very likely that the low 
metallicity of the VAS is only an effect of their highly active 
chromospheres, and not an indication of older age. A similar 
conclusion was recently reached by Favata et al. (1997) and Morale et 
al. (1996) based on photometric and spectroscopic studies of a 
sample of very active K type stars. 

\section{The origin of the activity strip}

It is now possible to make two hypotheses in order to understand 
the activity strip and other trends on the ${\rm [Fe/H]}\times
\log R'_{\rm HK}$ diagram:
\beginlist
\item (a) stars with lower metallicities should present higher apparent 
chromospheric indices $\log R'_{\rm HK}$, since the Ca lines are 
shallower in the spectra of metal-poor stars (see Figure~2). This 
could produce an anticorrelation between [Fe/H] and $\log R'_{\rm HK}$.
\item (b) the chromospheric activity affects the photometric index 
$m_1$ in such a way that the larger the CE level in a star, the more 
metal-poor this star would appear (Gim\'enez et al. 1991). This can 
also produce an anticorrelation as seen in the activity strip beyond
the Vaughan-Preston gap.
\endlist

The first hypothesis can account well for the deviations between 
isochrone and chromospheric ages seen amongst the IS (Figures 1a and b).
For the active stars, this is not as obvious, and 
Fig.~1 suggests that these stars require different corrections,
as compared to the inactive stars. If all active stars were close 
binaries of RS CVn or BY Dra types, this 
could also explain the activity strip. In that case, close binaries 
with ${\rm [Fe/H]} \ge +0.0$ would not be found amongst the VAS, 
but in the AS group. However, we have found no preferred region 
on the ${\rm [Fe/H]} \times \log R'_{\rm HK}$ diagram where the known 
resolved binaries (not only close binaries) in our Sample~B are located.
Moreover, it is well known that many AS stars are not binaries. 

The second hypothesis is particularly suitable to explain the activity
strip, since it does not require all active stars to be binaries. In 
fact, the shape of the $\log R'_{\rm HK}\times{\rm [Fe/H]}$ diagram, 
showing a quite abrupt change in the [Fe/H] distribution at the 
Vaughan-Preston gap, suggests that different processes are at work
on both sides of the gap. 

To test hypothesis (b), we have searched for spectroscopic 
metallicities in Cayrel de Strobel et al. (1997) for the stars in 
Sample~B. For the sake of consistency, we have used the most recent 
determination of [Fe/H] for stars having several entries 
in this catalogue. Note that the spectroscopic metallicity is not 
likely to be affected by the chromospheric activity, and could serve 
as a tool to check whether our photometrically derived [Fe/H] 
abundances are underestimated. 
\beginfigure{5}
  \vskip 6 cm
  \caption{{\bf Figure 5.} The difference between the spectroscopic 
   and photometric metallicity [Fe/H] as a function of the CE level.}
\endfigure
\beginfigure*{6}
  \vskip 12 cm
  \caption{{\bf Figure 6.} Simulation of the ${\rm [Fe/H]}\times
  \log R'_{\rm HK}$ diagram for a set of 1750 fictitious stars: a) 
   age--metallicity relation assuming a constant stellar birthrate; 
   b) CE--metallicity diagram assuming no metallicity dependence of 
   the $\log R'_{\rm HK}$ index; c) CE--metallicity diagram adopting 
   Eq. (4), and correcting the metallicity of the active stars 
   according to Eqs. (5); d) the same as panel c, but the metallicity 
   is corrected using Eq. (6).}
\endfigure
In Figure 5, we show the difference $\Delta{\rm [Fe/H]}$ between 
the spectroscopic and the photometric metallicities as a function of 
$\log R'_{\rm HK}$. As in the previous figures, two distinct trends 
can be seen here. For IS ($\log R'_{\rm HK} < -4.75$), the scatter in 
$\Delta{\rm [Fe/H]}$ is high  but the stars appear very well 
distributed around the expected value $\Delta{\rm [Fe/H]}\approx 0$. 
A gaussian fit to the $\Delta{\rm [Fe/H]}$ distribution for these
stars gives a mean $\langle\Delta\rangle \approx 0.03$ dex and
standard deviation $\sigma \approx 0.13$ dex. Such scatter is very likely 
the result of observational errors as well as of the inhomogeneity 
of the spectroscopic data, which was taken from several sources. For 
the AS ($\log R'_{\rm HK} > -4.75$), the data clearly shows that there 
is a {\it positive} $\Delta{\rm [Fe/H]}$, with few deviating points. 
According to a Kolmogorov-Smirnov test, the hypothesis that the
$\Delta{\rm [Fe/H]}$ distributions are the same for both inactive and
active stars can be rejected to a significance level lower than 0.05.

The paucity of points in this part of the diagram makes the relation 
between $\Delta{\rm [Fe/H]}$ and  $\log R'_{\rm HK}$ somewhat confusing. 
As a first aproximation, we will assume that $\Delta{\rm [Fe/H]}$ 
is independent of the CE level, so that it can be written as
$$\eqalignno{%
    \Delta{\rm [Fe/H]}& = 0.17,\qquad\log R'_{\rm HK} > -4.75 & \startsubeq\cr
    & = 0\phantom{.17},\qquad\log R'_{\rm HK} \le -4.75 &
    \stepsubeq\cr}$$
This correction has been applied to the active stars of Sample~B
to find the expected metallicity for these stars. The value 0.17 dex 
corresponds to the mean $\Delta{\rm [Fe/H]}$ of the active stars in 
Figure~5, excluding the four stars with negative $\Delta{\rm [Fe/H]}$. 
If these stars are included, the mean $\Delta{\rm [Fe/H]}$ decreases
down to about 0.11 dex.

From a more physical point of view, it is expected that 
$\Delta{\rm [Fe/H]}$ increases as the CE level increases 
(Gim\'enez et al 1991; Favata et al. 1997). Therefore, we have 
fitted a straight line  by eye again excluding the deviating stars, 
and assuming that $\Delta{\rm [Fe/H]}\approx 0$ for 
$\log R'_{\rm HK}\le -4.75$. The obtained expression for this 
approximation is 
$$\Delta{\rm [Fe/H]}=2.613+0.550\log R'_{\rm HK},\eqno\stepeq$$
as can be seen by the straight line in Fig.~5. 

To explain the shape of the ${\rm [Fe/H]}\times\log R'_{\rm HK}$ 
diagram, we have made a simulation of the diagram in Figure 3, 
using a set of 1750 fictitious stars. This set was 
built in a similar fashion to that described in Rocha-Pinto \& 
Maciel (1997), using a constant stellar birthrate and a cosmic 
scatter in [Fe/H] of 0.20 dex. In Figure 6a we show the 
age--metallicity relation for this set of stars. Note that this
figure is shown for illustrative purposes only, and we have made no 
attempts to use a relation closer to the real solar neighbourhood 
relation. This age--metallicity relation is transformed into a 
CE--[Fe/H] relation in Figure 6b, using Eq. (1). This would be 
the shape of the ${\rm [Fe/H]}\times
\log R'_{\rm HK}$ diagram if a CE--age relation and an 
age--metallicity relation exist, and if hypotheses (a) and (b) 
above were not valid. It can be seen that Fig.~6b does not reproduce 
the observed trends of the diagram of Figure~3. In Figure 6c, we 
show the same diagram assuming that the ages have an excess 
$\Delta(\log t)$ relative to the chromospheric ages, as given in 
Eq.~(4), and that the metallicities of the AS and VAS have a 
deficiency of $\Delta{\rm [Fe/H]}$ as given by Eqs. (5). Figure~6d 
is similar to Figure 6c, except that Eq. (6) is used instead of 
Eqs.~(5). In Figures 6b, c and d, the vertical dashed lines separate 
the four populations according to CE level, as in Figure~3. 
Figure~6c shows a clear progress in reproducing Fig.~3 as compared
with Fig.~6b, and Fig.~6d can in fact reproduce fairly well the 
trends of the observed CE--metallicity diagram of Figure~3. Note 
that all old metal-poor stars in Figure~6b look more active, that is, 
younger, in Figure~6d. Also, some young metal-rich stars are shifted 
to the VIS group in agreement with the observations. The activity 
strip beyond the Vaughan-Preston gap is also very well reproduced 
in Fig.~6d. Note that Figure~6c does not reproduce the activity 
strip, suggesting that Eq. (6) is physically more meaningful than 
Eqs.~(5).

Unfortunately, there are no calibrations for stellar parameters 
in chromospherically active stars, which would make possible to
check whether  Eq. (6) gives the right corrections. However, the good 
agreement between our simulations and Figure~3 suggests that our 
procedure to correct the photometric [Fe/H] is fairly satisfying. 

Another illustration of the correction procedure proposed in this 
paper is presented in Fig. 7, where we show the uncorrected 
age-metallicity relation (AMR) for Sample B (Fig. 7a), as well as
the obtained relation after our correction procedure has been applied
(Fig. 7b). Also shown are four average relations from the literature.
It can be seen that the uncorrected sample presents a very large
scatter, so that stars of all metallicities can be found at
almost all ages. On the other hand, the corrected sample shows a
much better agreement with the tendency presented by all previous
relations, yet preserving some scatter averaging 0.3 dex as observed.
A full discussion of this age-metallicity relation will be the subject
of a forthcoming paper.
\beginfigure{7}
  \vskip 12 cm
  \caption{{\bf Figure 7.} a) Uncorrected and b) corrected age-metallicity
  relation for Sample B. Also shown are some AMR from the literature.}
\endfigure
\section{The CE--age relation for active stars}

Although a CE--age relation exists amongst late-type dwarfs, its shape
is not well defined. The main uncertainty seems to occur for stars 
beyond the Vaughan-Preston gap. Soderblom et al. (1991) have 
shown that the slope of the CE--age relation could be steeper for 
$\log R'_{\rm HK}>-4.75$. This would dramatically affect the 
interpretation of the peaks in the chromospheric age distributions 
as a result of star formation bursts in our Galaxy 
(Barry 1988; Noh \& Scalo 1990; Rocha-Pinto \& Maciel 1997).
\beginfigure{8}
  \vskip 6 cm
  \caption{{\bf Figure 8.} $\delta c_1\times\log R'_{\rm HK}$ diagram 
    for Sample B. The $\delta c_1$ index can be regarded as a rough 
    age indicator for our sample.}
\endfigure
The calibration of the CE--age relation for the active 
stars, using $T_{\rm eff}$ and $M_V$ derived from photometry, is not 
as simple as for the inactive stars, since we do not know very well 
to what extent the photometric indices are affected by the 
chromospheric activity. Consider the example shown in Figure~8.
In this plot we show a $\delta c_1\times\log R'_{\rm HK}$ diagram 
for all stars in Sample B. The $\delta c_1$ index is a measure of the 
distance of the star from the ZAMS. In a sample with a narrow 
$T_{\rm eff}$ range, as is the case of a sample comprised only by 
G dwarfs, it is possible to roughly associate a high $\delta c_1$ 
with a greater age (lower $\log R'_{\rm HK}$), and vice-versa. If 
the AS are really very young they should have 
$\langle\delta c_1\rangle \la 0.0$, and $\delta c_1$ should become 
higher as lower $\log R'_{\rm HK}$ values are considered. From Fig.~8,
despite the scatter, we can see a clear relation between 
$\delta c_1$ and $\log R'_{\rm HK}$ for the IS and probably also 
the VIS, as expected on the basis of the previous discussion. 
However, the AS group does not show $\delta c_1$ indices 
typical of a very young population, and there seems to be no 
continuity in the $\delta c_1\times\log R'_{\rm HK}$ relation  for 
inactive and active stars. The plot even suggests that the active 
stars could be composed by stars with a variety of ages. 

According to Gim\'enez et al. (1991) and Favata et al. (1997), the 
$c_1$ index is less likely to be affected by the chromospheric 
activity than the $m_1$ index. However, in the calculation of 
$\delta c_1$ we have also made use of the stellar metallicity, to 
account for the [Fe/H] dependence of the ZAMS. This is probably the 
main reason for not finding $\langle\delta c_1\rangle \sim 0.0$ for 
the active stars. If $\delta c_1$ is incorrect, then $\Delta M_V$ 
would also be miscalculated, and the isochrone ages found 
for these stars would give no information about the real stellar age. 
Moreover, the photometric metallicity of the star is needed to 
interpolate between the isochrones of different metallicities, which 
would be another source of errors. This would produce a systematic 
error towards greater ages for the active stars. 

In view of these considerations, it is not surprising that most active 
stars deviate from the relation between age excess and [Fe/H] in 
Figures 1a and 1c. Note that all single stars with 
$\log R'_{\rm HK} > -4.75$ in Figure~8 from Soderblom et al. (1991) 
also have isochrone ages (ranging nearly from 2 to 4 Gyr) 
systematically greater than their ages from the calibration, 
which could be caused by some of these effects.
Until we have photometric calibrations for the stellar parameters in 
active stars, we cannot use photometrically derived temperatures and 
magnitudes to obtain stellar ages. The best way to calibrate the 
CE--age relation beyond the Vaughan-Preston gap would be to use only 
open clusters. These objects could have ages determined by 
isochrone fitting, a procedure which is not affected by the CE level 
of the cluster stars. However, these clusters are generally distant 
and their low mass stars are quite faint to allow CE measurements 
(Soderblom et al. 1991). On the other hand, a sample of stars 
having $T_{\rm eff}$ and [Fe/H] calculated from their spectra, and 
$M_V$ from their parallaxes, could have accurate isochrone ages. 
This is what happens in Figure 1b, where $M_V$ is calculated from 
trigonometrical parallaxes. In this figure, the active stars follow 
more closely the age excess relation given by Equation (4).
\begintable*{1}
\caption{{\bf Table 2.} -- Spatial velocities for VAS and AS in Sample B}
{\halign{%
\hfil \rm # \hfil &  \quad \hfil $#$ & \quad \hfil $#$ & \quad \hfil$#$ & \quad 
\hfil $#$ & \quad \hfil \rm # \hfil &  \quad \hfil $#$ & \quad \hfil $#$ & \quad 
\hfil$#$ & \quad \hfil $#$ & \quad \hfil \rm # \hfil &  \quad \hfil $#$ & \quad 
\hfil $#$ & \quad \hfil$#$ & \quad \hfil $#$  \cr
HD & \log R'_{\rm HK} & U & V & W & HD & \log R'_{\rm HK} & U & V & W & HD & 
\log R'_{\rm HK} & U & V & W \cr
\noalign{\vskip 10 pt}
166 & -4.33 & -15 & -23 & -10 & 61994 & -4.59 & 2 & -23 & -20 & 144872 & -4.74 & 55 & 5 & 5  \cr
1835 A & -4.44 & -33 & -13 & 0 & 65721 & -4.67 & -44 & -9 & -26 & 147513 A & -4.52 & 17 & -3 & -1 \cr
4391 & -4.55 & -22 & 0 & 7 & 72905 & -4.33 & 10 & 0 & -10 & 150706 & -4.59 & 20 & -4 & -13 \cr
10780 & -4.61 & -23 & -15 & -4 & 74576 & -4.31 & -26 & -9 & -1 & 152391 & -4.39 & 84 & -108 & 10 \cr
10800 & -4.60 & -11 & -6 & -10 & 74842 & -4.58 & 27 & -9 & -21 & 160346 & -4.71 & 19 & -1 & 12 \cr
11131 B & -4.47 & 20 & 3 & -5 & 82106 & -4.43 & -36 & -13 & 0 & 165185 & -4.43 & 13 & 5 & -9 \cr
13445 & -4.74 & -109 & -80 & -24 & 88742 & -4.69 & -31 & -44 & -1 & 165401 & -4.65 & -83 & -84 & -38 \cr
17051 & -4.65 & -27 & -15 & -8 & 88746 A & -4.70 & -44 & -6 & -1 & 181321 & -4.31 & -13 & -5 & -7 \cr
17925 & -4.30 & -15 & -16 & -11 & 101501 & -4.50 & 7 & -15 & -4 & 189087 & -4.63 & -34 & -14 & 5 \cr
20630 & -4.45 & -22 & -4 & -4 & 103431 & -4.68 & -70 & -38 & -12 & 189931 & -4.64 & -44 & -63 & -5 \cr
20766 & -4.65 & -63 & -42 & 13 & 110010 & -4.47 & -8 & -21 & -13 & 196850 & -4.64 & -1 & -23 & -34 \cr
22049 & -4.47 & -3 & 8 & -20 & 110833 & -4.72 & -21 & -24 & 13 & 202628 & -4.73 & -11 & 1 & -27 \cr
25680 & -4.54 & -25 & -15 & -7 & 115043 & -4.43 & 15 & 3 & -8 & 203244 & -4.39 & 1 & 9 & -23 \cr
30495 & -4.54 & -23 & -9 & -3 & 120237 A & -4.75 & -49 & -60 & -5 & 205905 & -4.55 & -40 & -15 & -15 \cr
36435 & -4.44 & 12 & -3 & -19 & 124580 & -4.54 & 11 & 0 & -16 & 206860 & -4.42 & -13 & -20 & -8 \cr
37394 A & -4.43 & -12 & -21 & -13 & 128165 & -4.74 & -21 & 7 & 11 & 209100 & -4.56 & -78 & -41 & 3 \cr
41700 C & -4.35 & -35 & -11 & -15 & 129333 & -4.23 & -7 & -35 & -3 & 216803 & -4.27 & -6 & -9 & -11 \cr
42807 & -4.44 & 3 & -25 & -6 & 130948 & -4.45 & 3 & 7 & -6 & 219709 A & -4.62 & -18 & -40 & -15 \cr
53143 & -4.52 & -24 & -17 & -15 & 131582 A & -4.73 & -73 & -71 & 18 & 220182 & -4.41 & -56 & -15 & -2 \cr
58830 & -4.49 & 5 & 15 & 21 & 131977 & -4.49 & 46 & -21 & -31 & 222335 & -4.73 & -2 & -39 & 6 \cr
59967 & -4.36 & -11 & -4 & -7 & 134319 & -4.33 & -33 & -15 & -1 &  &  &  &  & \cr
}}
\endtable
A mean age for a group of stars can also be determined on the basis 
of its kinematical properties. Soderblom (1990) has investigated 
the kinematics of about 30 AS, and has found a mean age of 1 to 2 Gyr 
for the group. Nevertheless, he has found some active stars with 
kinematics of old population stars. We decided to reinvestigate the 
kinematics of the active stars, using a larger data sample. We have 
searched for spatial velocities for each  active star of Sample B in 
Gliese \& Jahrei\ss\ (1991) and Soderblom (1990). We have also 
calculated spatial velocities for 17 stars, from their radial and 
tangential velocities, proper motions and parallaxes, obtained from
the SIMBAD database, using the formulae from Johnson \& Soderblom (1987). 
Table 2 lists 62 active stars with U, V and W velocities (km/s). 
\beginfigure{9}
  \vskip 12 cm
  \caption{{\bf Figure 9.} Spatial motions of 62 active stars
   listed in Table 2.}
\endfigure
\beginfigure{10}
  \vskip 6 cm
  \caption{{\bf Figure 10.} Velocity dispersion $\sigma_W$ as a 
  function of $\log R'_{\rm HK}$. The triangles show the 
  $\sigma_W$--age relation (Wielen 1977), and we have used the 
  CE--age calibration by Soderblom et al. (1991) 
  to find the $\log R'_{\rm HK}$ index for each age.}
\endfigure
Figure 9 shows the spatial motions of these stars. These panels are 
similar to those in Soderblom (1990). However, we have found several 
stars with kinematics of old and intermediate-age population. Some 
of these stars have $\log R'_{\rm HK}\sim -4.75$, and could be 
inactive stars. The velocity dispersions $\sigma_U$, $\sigma_V$ and 
$\sigma_W$ are 36, 30 and 14 km/s, respectively, which indicate
a mean age of 2 to 4 Gyr for the AS group according to  
Wielen (1977). However, an age of 1 to 2 Gyr would be found, in 
agreement with Soderblom (1990), if we disregard the stars with 
large velocities. It is still early to theorize upon the nature 
of such high-velocity active stars, and a larger sample is 
needed to allow firm conclusions.

We have also found a correlation between $\sigma_W$ and 
$\log R'_{\rm HK}$, which confirms the existence of a CE--age 
relation beyond the Vaughan-Preston gap. That is illustrated in 
Figure~10. In this figure, the points with error bars are the 
calculated $\sigma_W$ for 3 selected bins of our sample of active 
stars, and the triangles show the empirical $\sigma_W$--age relation 
according to Wielen (1977). The numbers near the triangles indicate 
a mean age for the stellar group with the corresponding $\sigma_W$. 
This relation was transformed into a $\sigma_W$--$\log R'_{\rm HK}$ 
relation using the CE--age calibration by Soderblom et al. (1991). 
Note that the velocity dispersions for the active stars are systematically 
larger than the dispersions of their presumed coeval stars. This result 
suggests that the slope of the CE--age relation for the active stars 
is flatter than the slope in Eq. (3) of Soderblom et al., contrary to 
the relation that they propose to preserve the constancy of the star 
formation rate. However, this conclusion must be regarded with 
caution, as it is not founded on a large database. In fact, 
considering only the open clusters in Figure~8 from Soderblom et 
al., we would find a steeper slope. Much work is needed to understand 
the behaviour of the CE--age relation for active stars. 

A program to measure spectroscopically $T_{\rm eff}$, [Fe/H], and 
abundance ratios in active stars is currently underway. Together 
with the new Hipparcos parallaxes, these data can cast some light on 
the discussion of the ages of the AS, and on the shape of the CE--age 
relation beyond the Vaughan-Preston gap.

\section*{Acknowledgements}

HJR-P thanks Gustavo F. Porto de Mello for an early suggestion of the 
relation between CE and age that has led to this work. We are indebted
to Dr. D. Soderblom for some comments on an earlier version of this
paper. This research has made use of the SIMBAD database, operated at 
CDS, Strasbourg, France. This work was supported by FAPESP and CNPq.

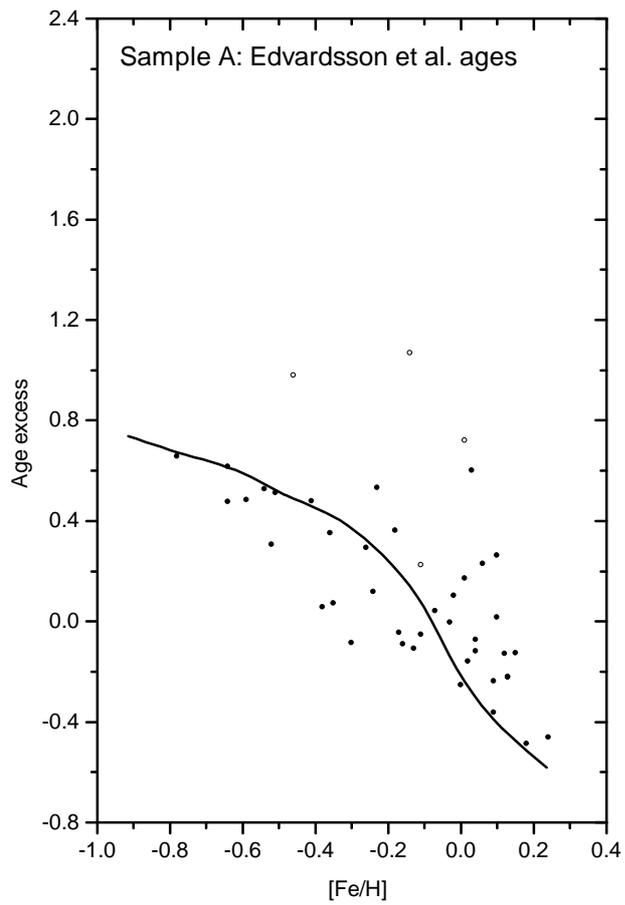
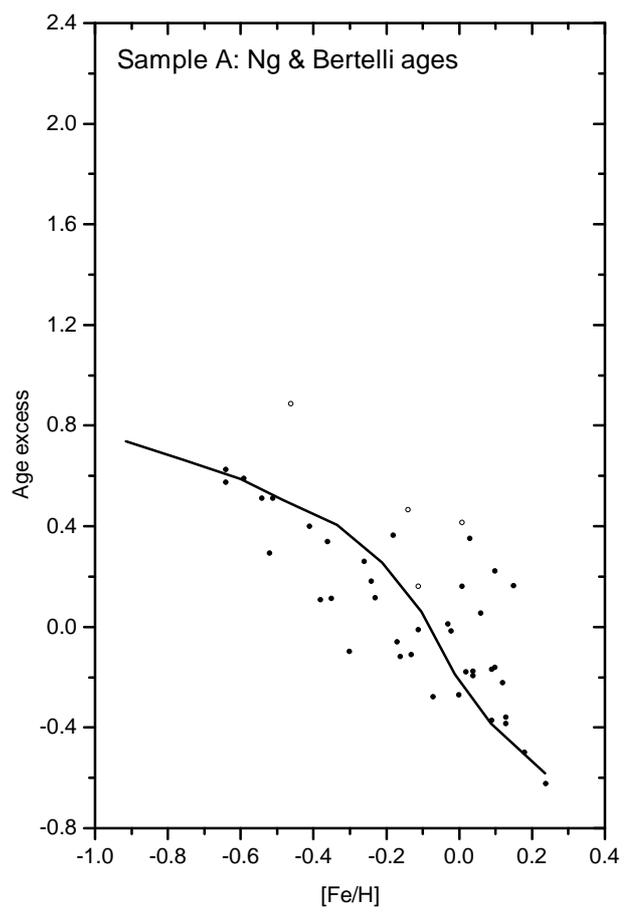
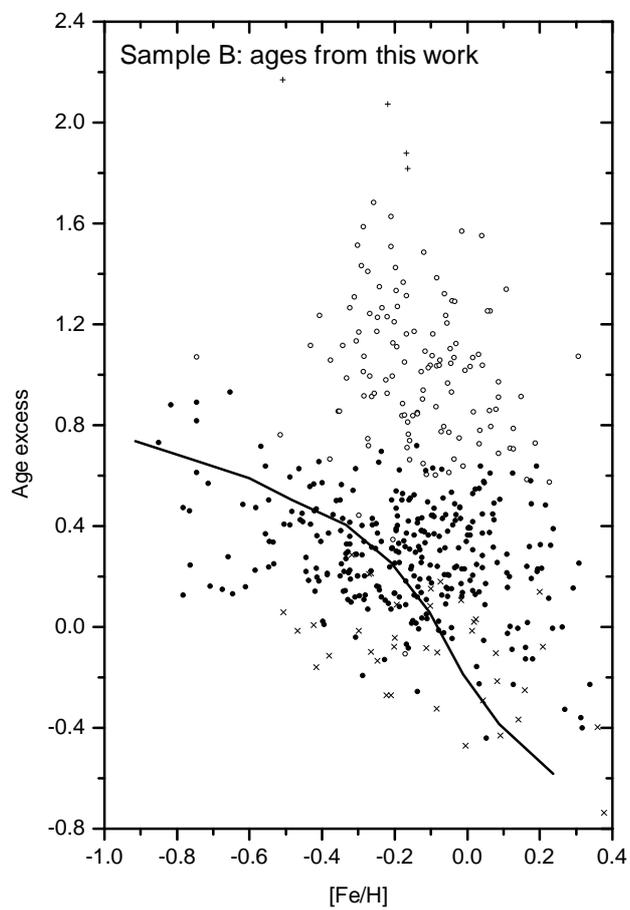
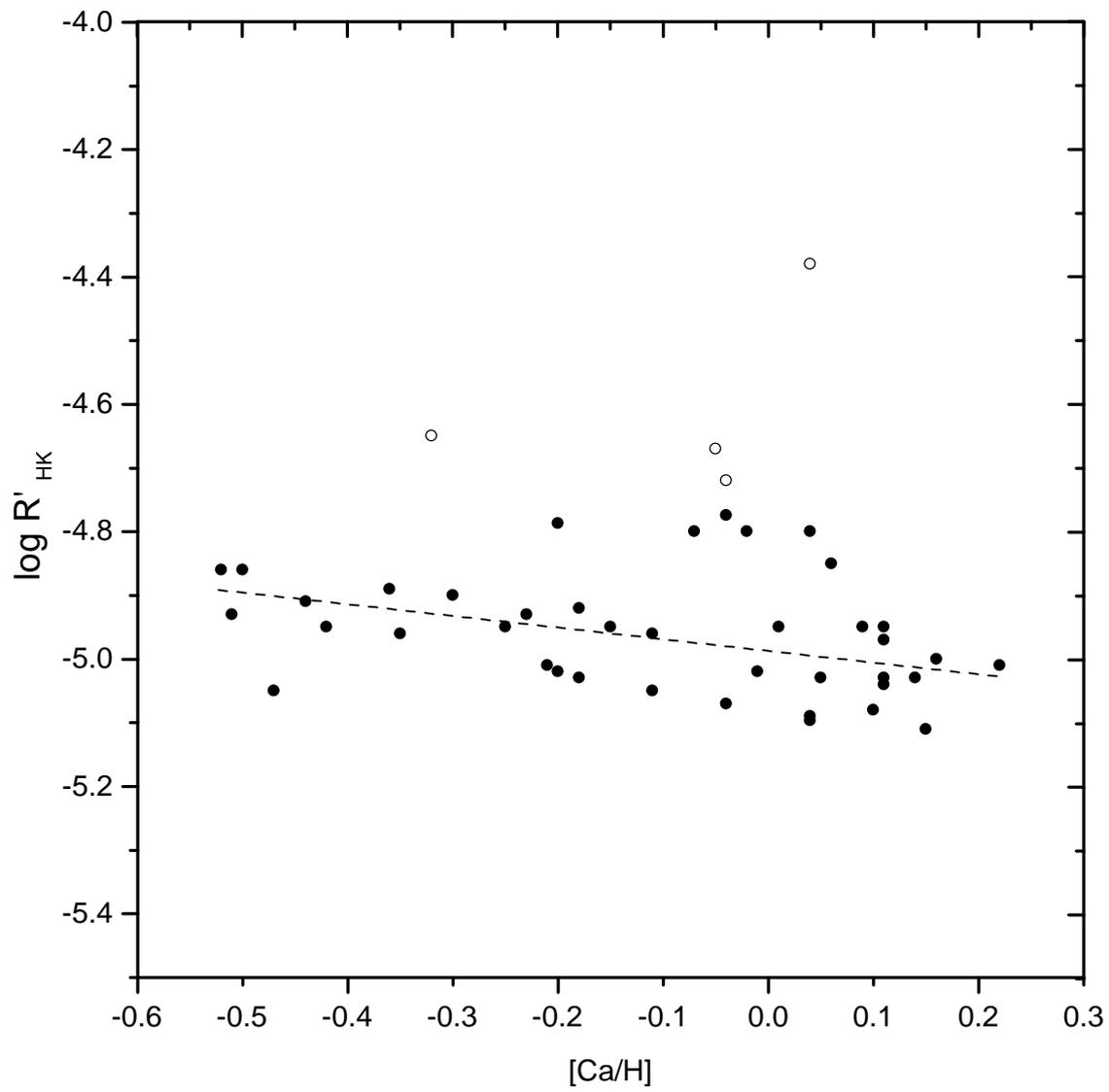
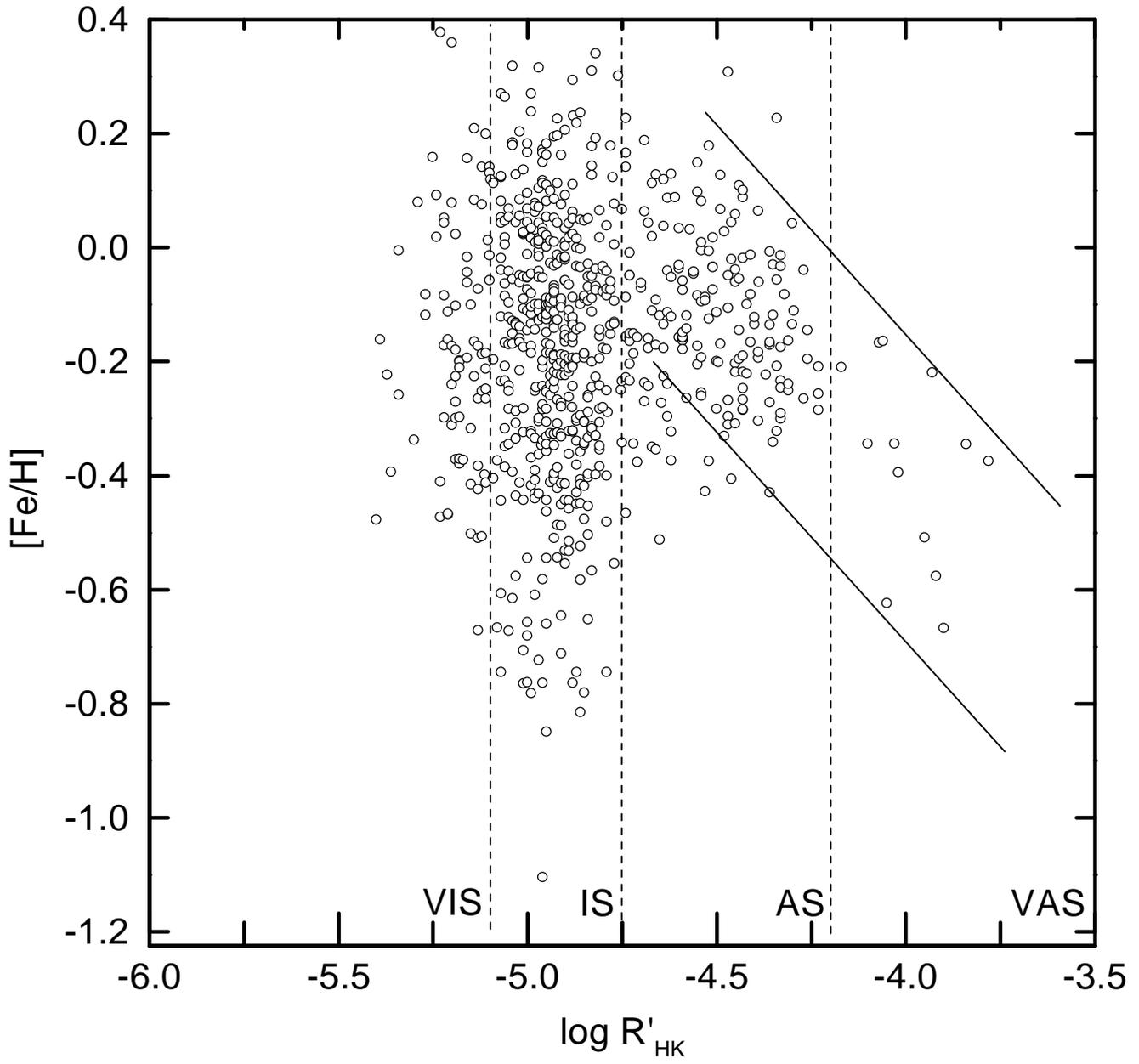
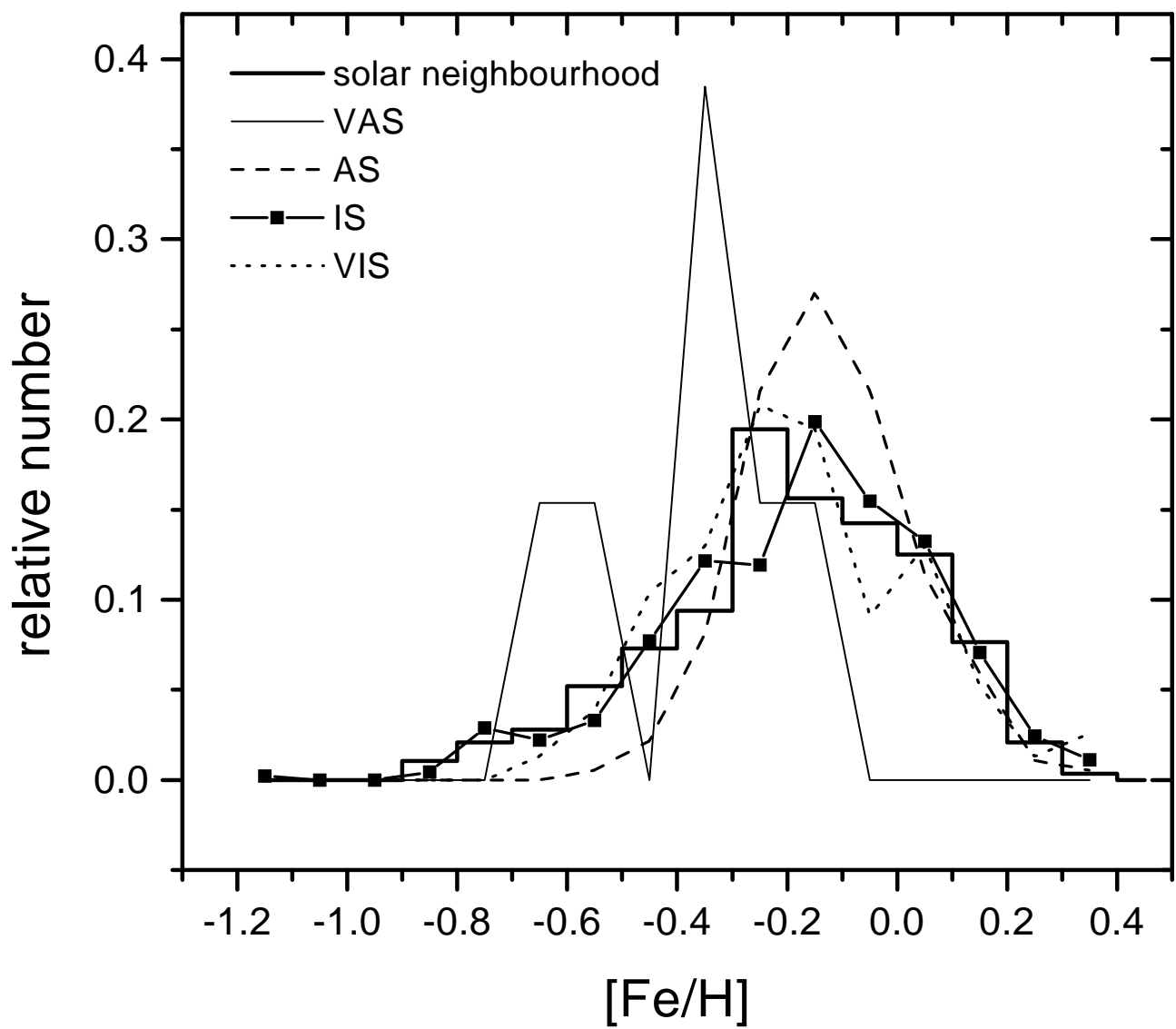
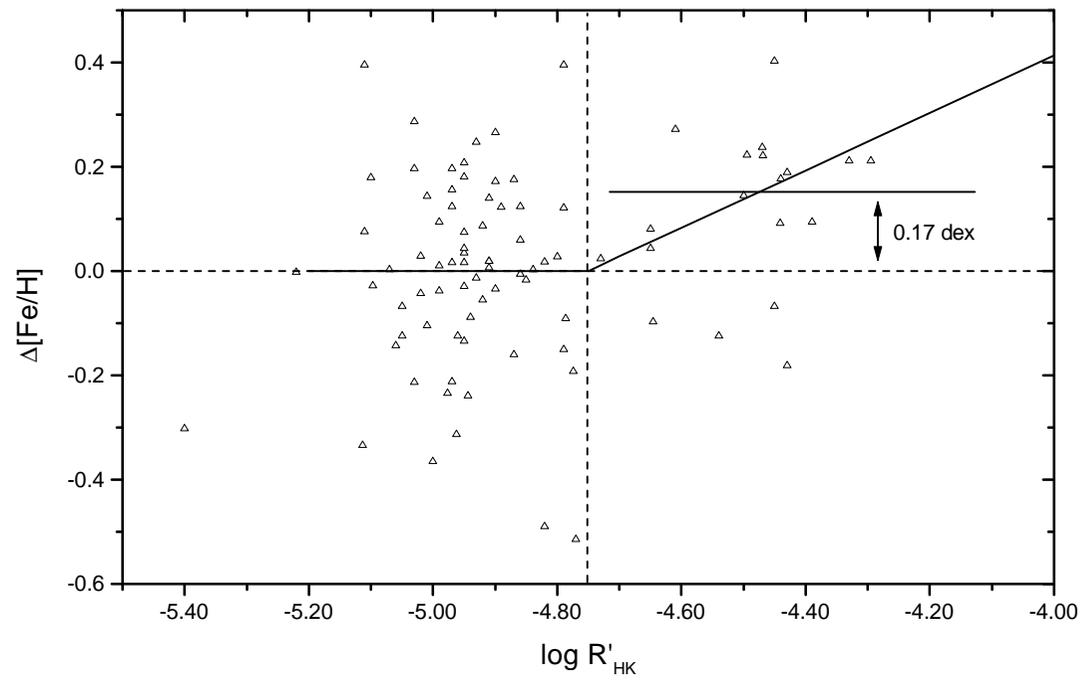
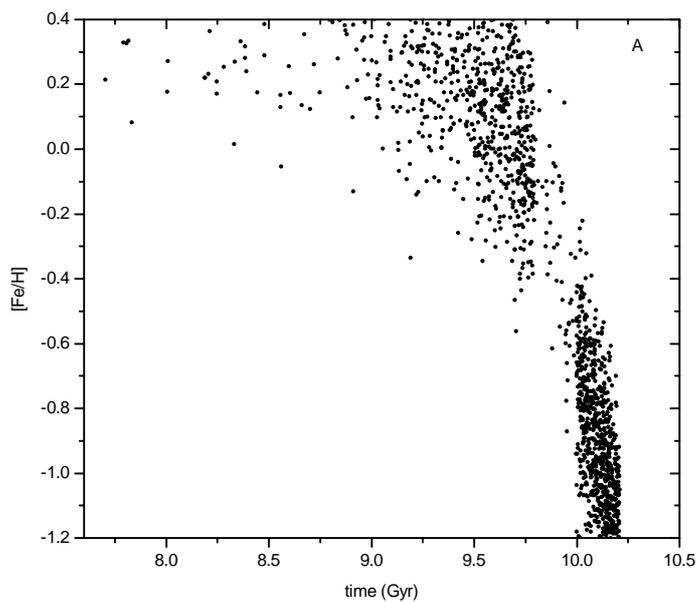
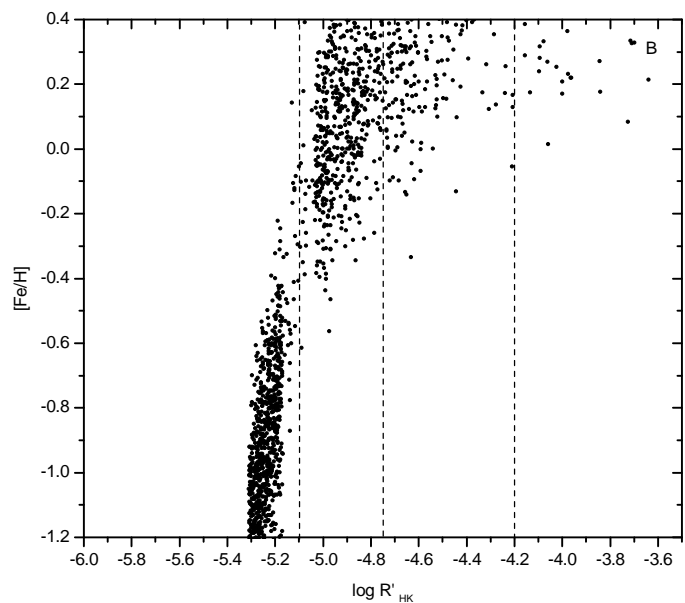
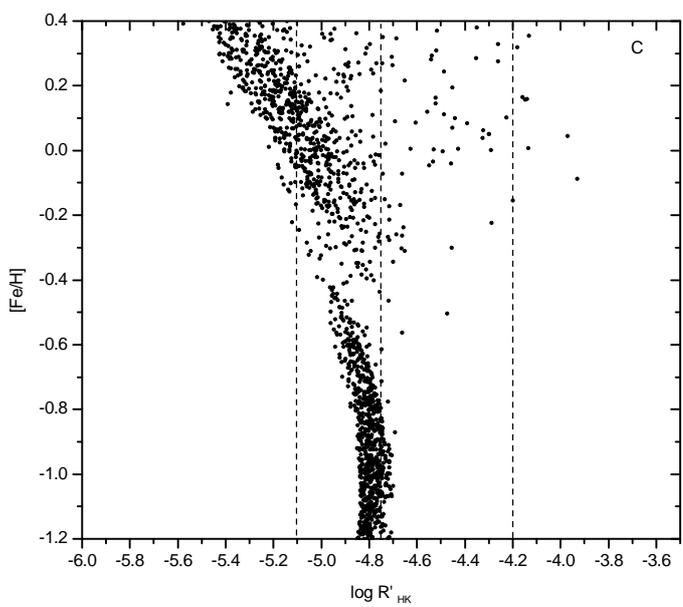
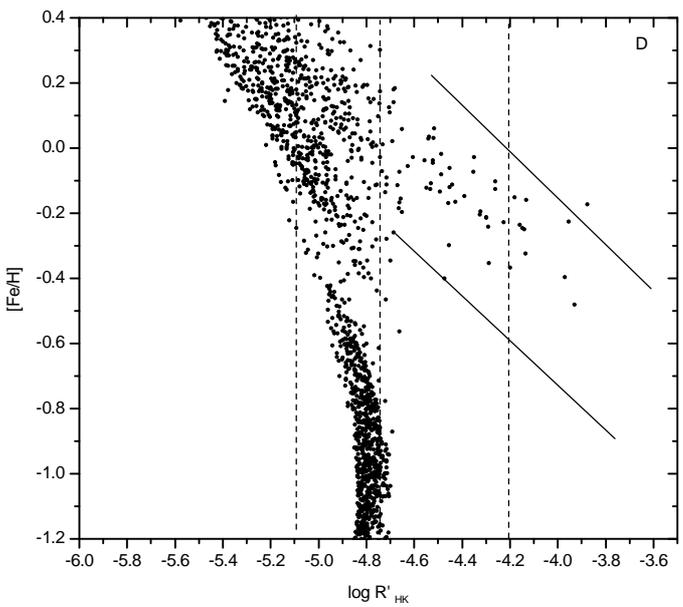
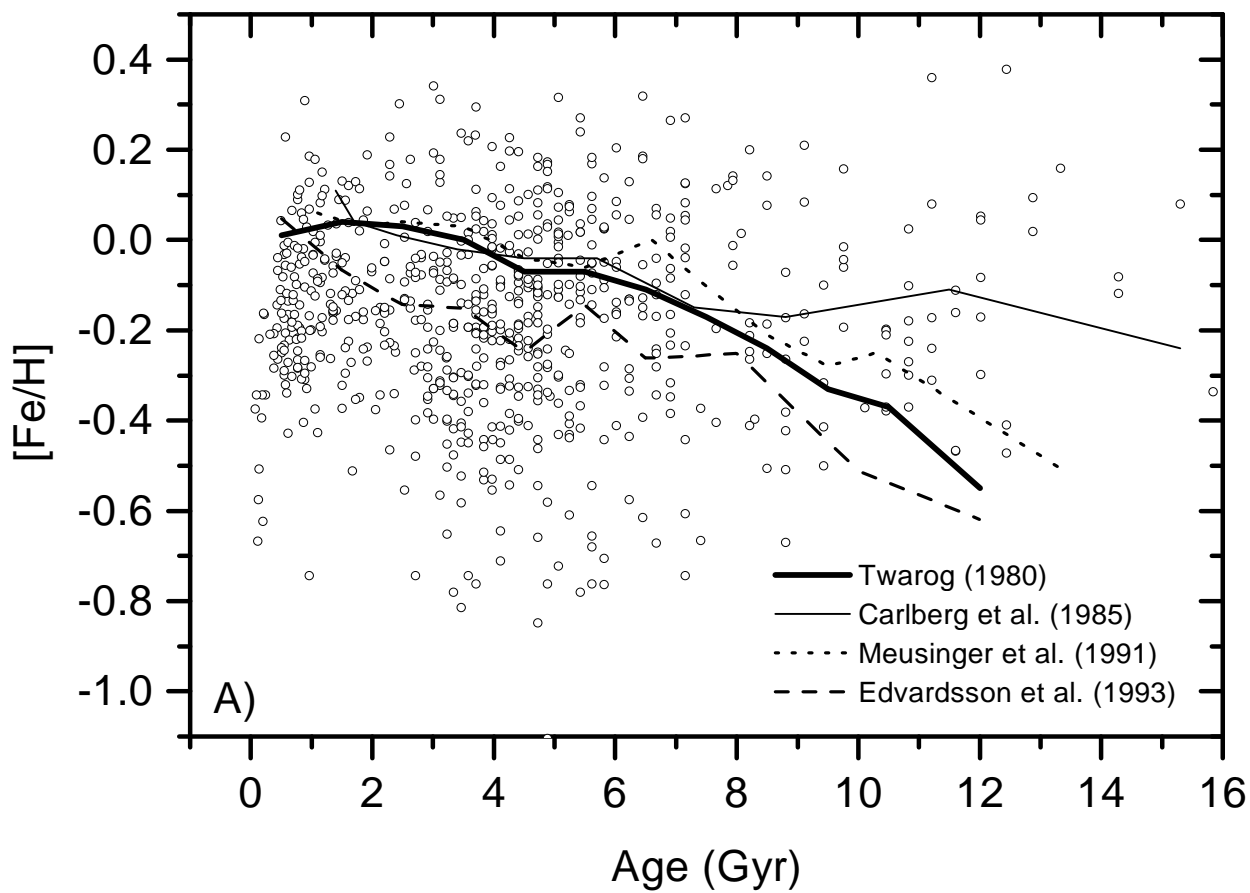
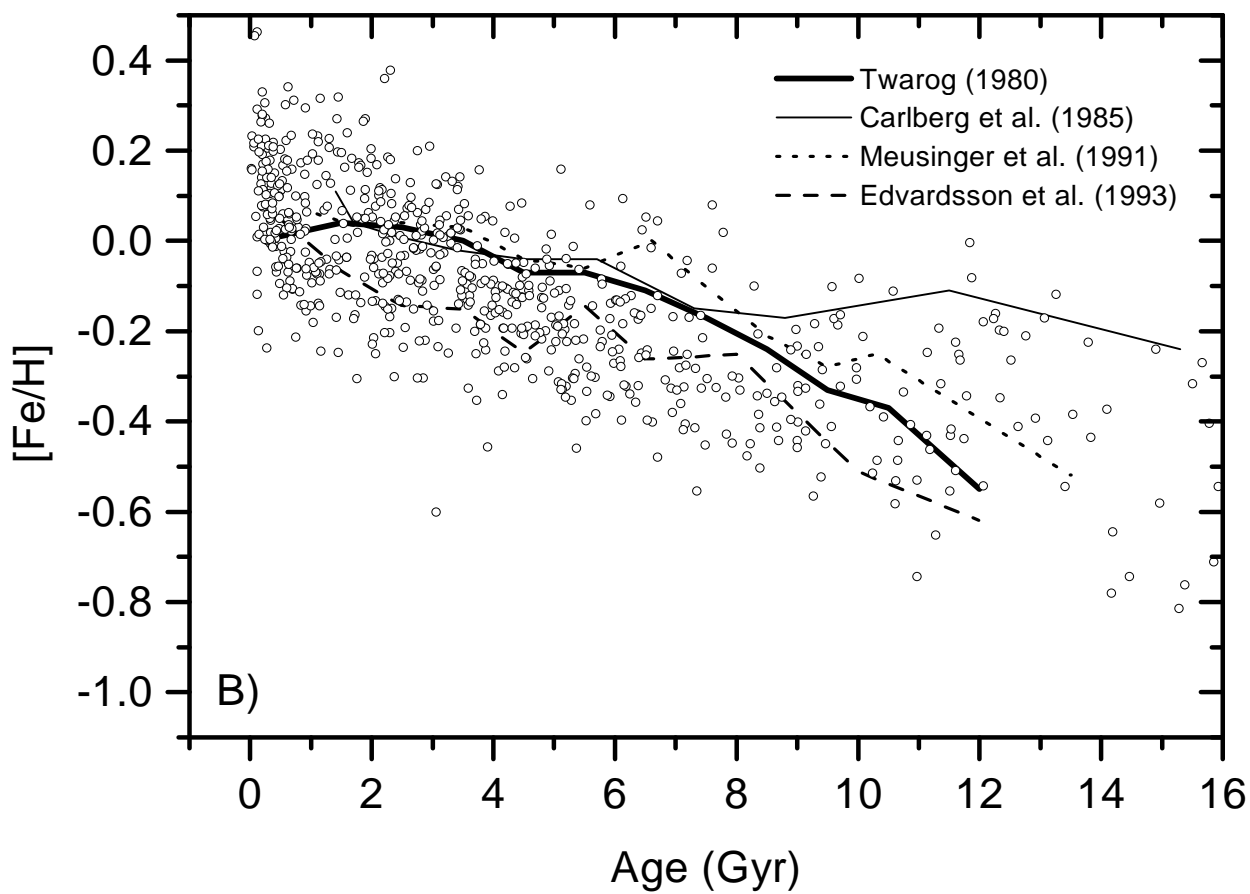
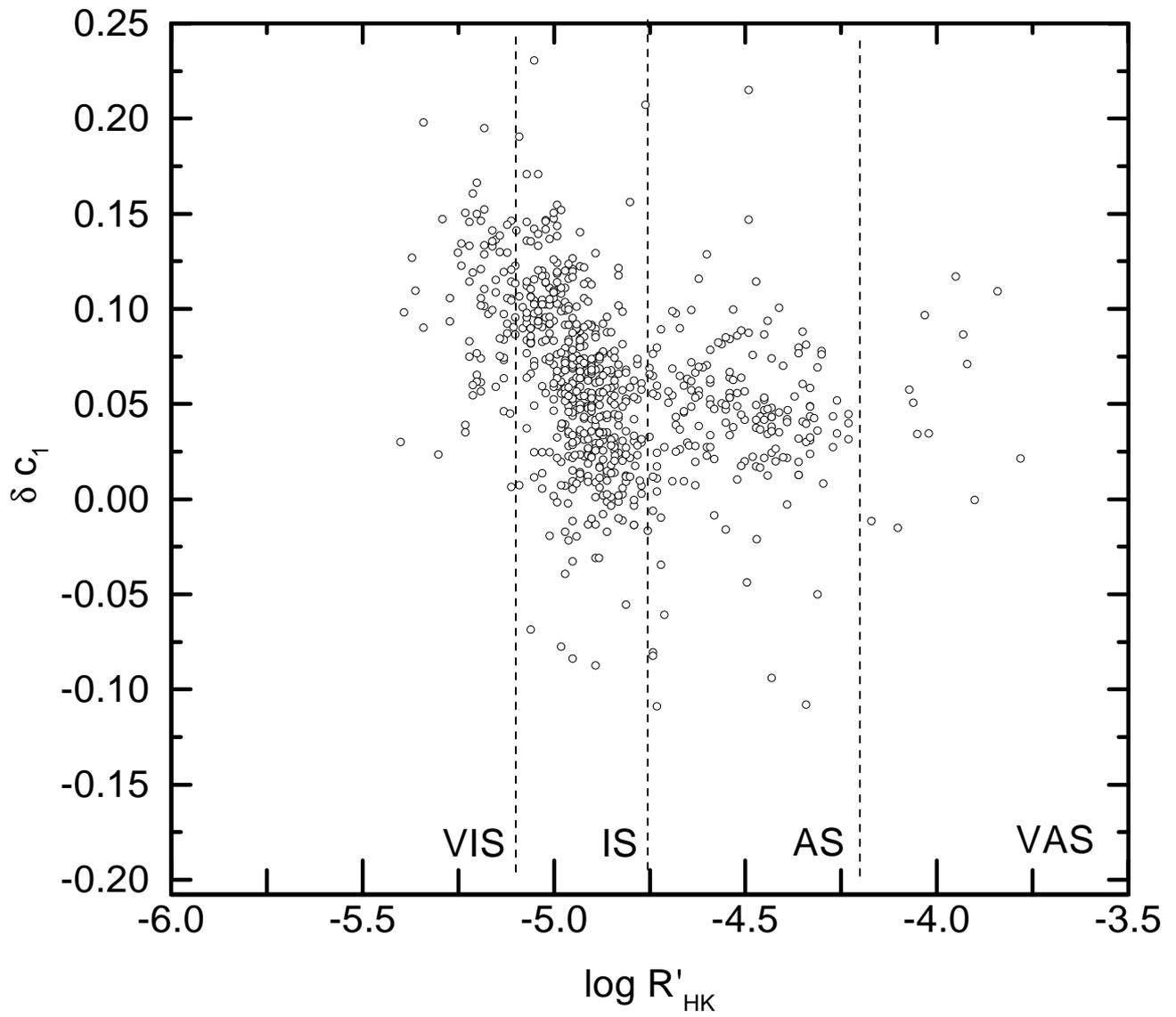
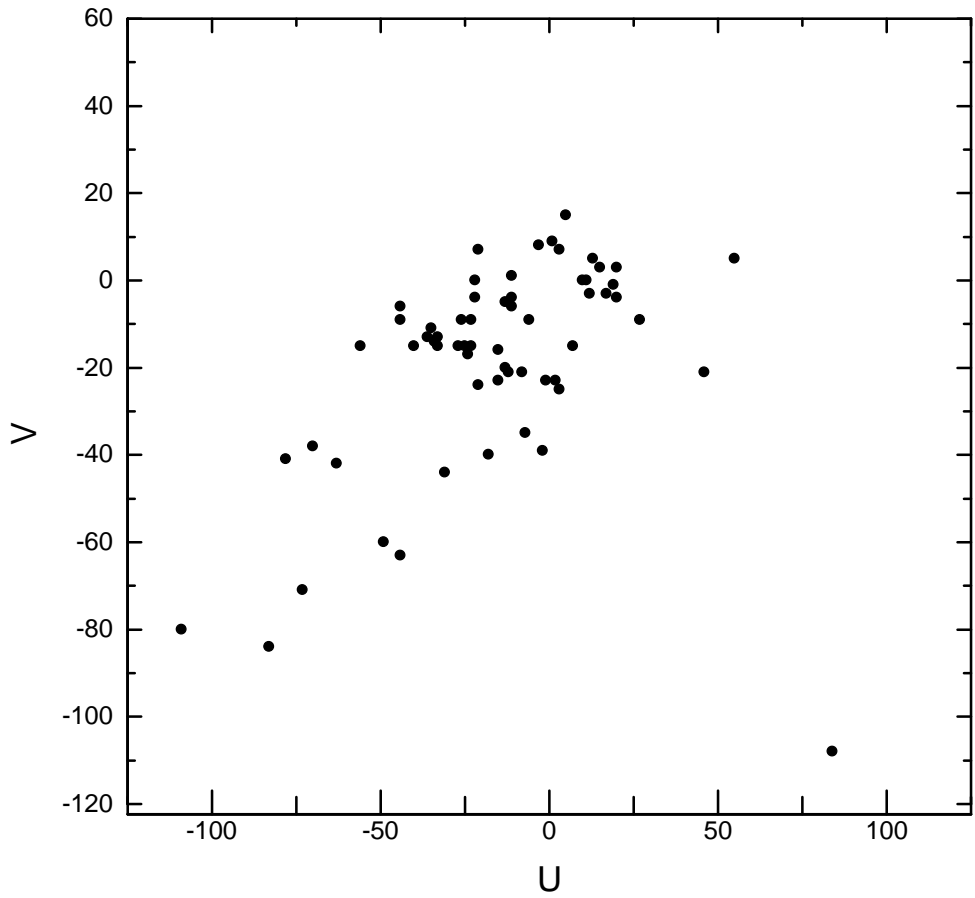
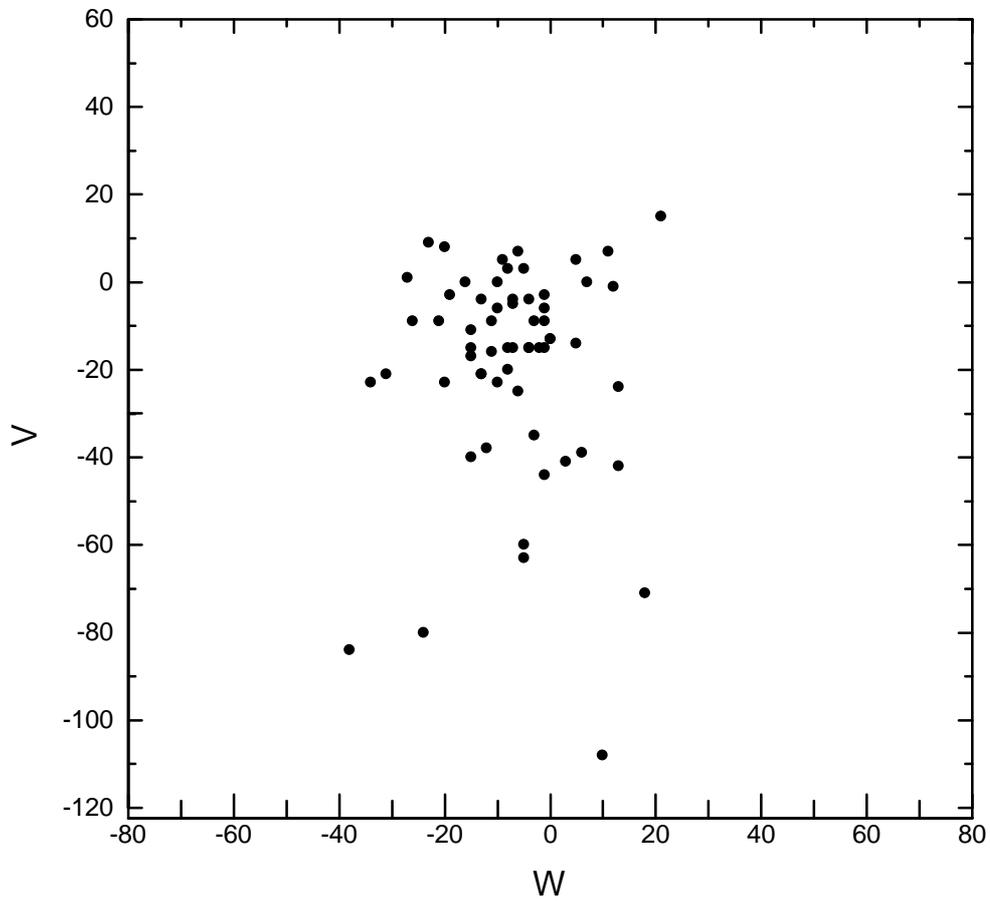
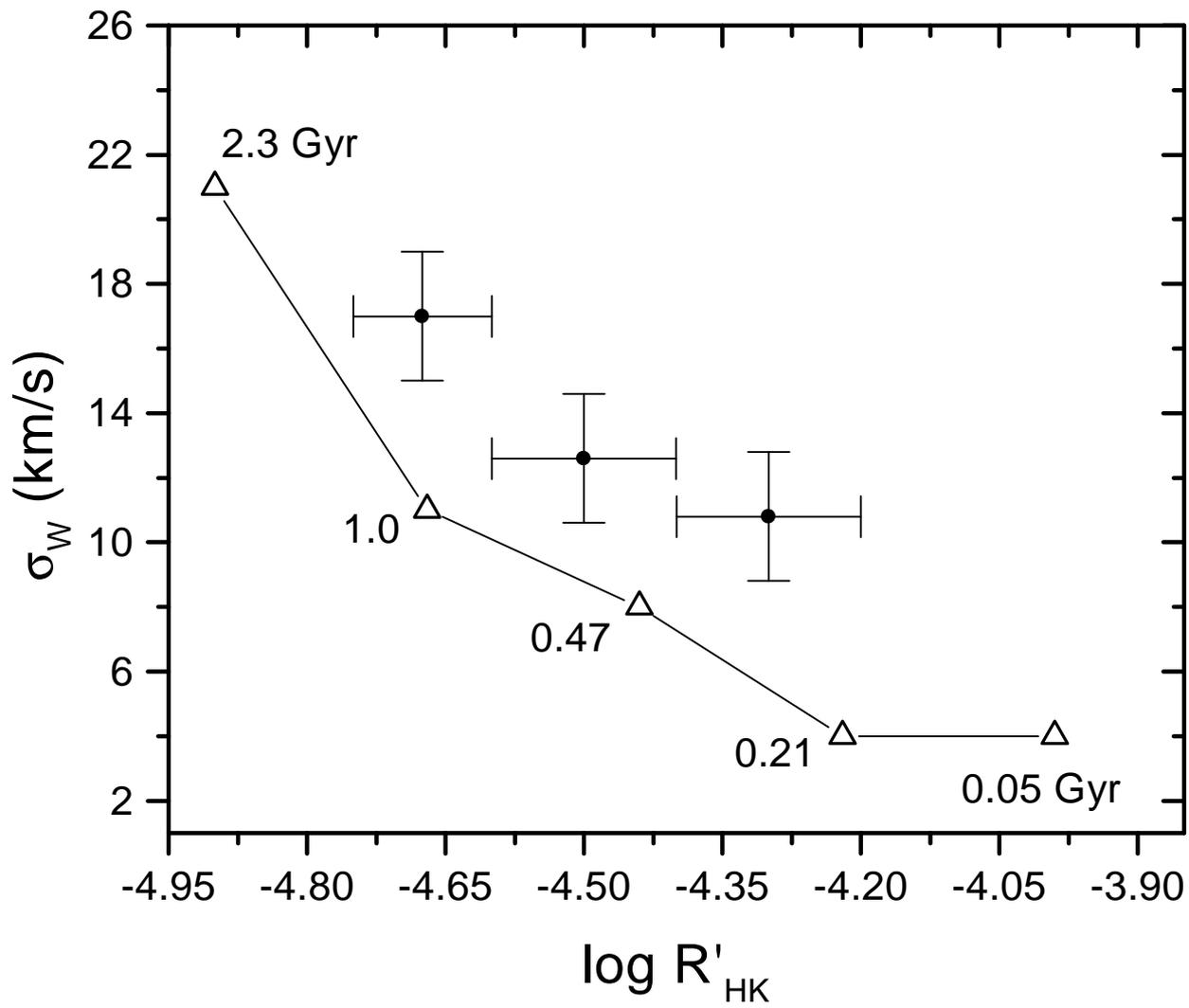
\bye